# Reversible Video Data Hiding Using Zero QDCT Coefficient-Pairs

Yi Chen[1] • Hongxia Wang[2] •Hanzhou Wu[3] • Yong Liu[1]

**Abstract**: H.264/Advanced Video Coding (AVC) is one of the most commonly used video compression standard currently. In this paper, we propose a Reversible Data Hiding (RDH) method based on H.264/AVC videos. In the proposed method, the macroblocks with intra-frame 4×4 prediction modes in intra frames are first selected as embeddable blocks. Then, the last zero Quantized Discrete Cosine Transform (QDCT) coefficients in all 4×4 blocks of the embeddable macroblocks are paired. In the following, a modification mapping rule based on making full use of modification directions are given. Finally, each zero coefficient-pair is changed by combining the given mapping rule with the to-be-embedded information bits. Since most of last QDCT coefficients in all 4×4 blocks are zero and they are located in high frequency area. Therefore, the proposed method can obtain high embedding capacity and low distortion.

✉Yi Chen
  yichen.research@gmail.com

✉Hongxia Wang
  hxwang@scu.edu.cn

Hanzhou Wu
  wuhanzhou_2007@126.com

Yong Liu
  liuyongresearch@163.com

1   School of Information Science and Technology, Southwest Jiaotong University, Chengdu 611756, P.R. China

2   College of Cybersecurity, Sichuan University, Chengdu 610064, P.R. China

3   Institute of Automation, Chinese Academy of Sciences (CAS), Beijing 100190, P. R. China

# 1 Introduction

Encryption and Data Hiding (DH) are two kinds of techniques to keep data secure. Their differences are that the former can hide the content of information and the latter can hide not only the content but also the existence of the information. Up to now, there are many encryption methods that have been proposed. For instance, Alassaf et al. proposed an improved SIMON cryptography for reducing the encryption time and keeping the practical trade-off between performance and security in [1]. Furthermore, to keep data securer, secret sharing technique [2-4] is proposed. Secret sharing technique can divide secret into many parts and each participator in secret sharing only manages his own part. Anyone cannot restore secret information individually unless the number of participators in restoring meets the requirement of secret sharing technique. For example, Gutub et al. presented a secret sharing scheme, which is based on counting, for multimedia applications in [4]. However, encryption technique can only hide the content but not the existence of information. Therefore, transmitting the encrypted data may be found that makes the encrypted data destroyed and stops the transmission. DH can be used to solve this problem because it can hide not only the content but also the existence of transmitting information.

DH is the science and art of imperceptibly hiding some secret data such as marks or messages, into a host media such as text, speech, image and video. According to the type of host media, DH is further classified into four types, i.e., text DH [5], speech DH [6], image DH [7-13], and video DH [14-18]. For example, in [5] Alsaidi et al. took the sensitive text and made use of the LZW compression algorithm to compress it. In the following, they exploited Advanced Encryption Standard (AES) to encrypt the compressed text to propose a text DH method. [5]. Liu et al. proposed a speech-based DH method in [6] and Zhang et al. proposed an image DH method by exploiting modification direction (EMD) in [13]. In addition, Ma et al. proposed a video DH scheme without intra-frame distortion drift in [15]. It can be said that different DH methods are applied in various application scenarios because of the certain requirements of people. However, DH cannot satisfy some specific application scenarios like medical image sharing, law forensics, multimedia archive management and military image protection. In these scenarios, the original cover should be able to be reconstructed. This requirement creates a novel research domain called Reversible Data Hiding (RDH). Likewise, RDH is in detail divided into three types, including speech RDH, image RDH and video RDH. In the past two decades, RDH has attracted a great deal of research interests [19]. Therefore, this has resulted in rapid development of RDH and makes RDH successfully exploited for digital watermarking [20-21], content authentication, privacy protection [22-26] and so on. Currently, although there are some RDH schemes in videos that have been reported. For instance, based on the work done by Ma et al. [15], Liu et al. utilized the secret sharing scheme [2-4] to propose a reversible video data hiding scheme with robustness and low bitrate growth in [27]. Moreover, Kim et al selected four middle frequency coefficients of 4×4 blocks to combine with Difference Expansion (DE) [19-21] to present a RDH method for H.264 bitstreams in [28]. However, it is not enough since the development of RDH mainly focuses on speech and images. Along with the development of video compression standards, such as H.264/Advanced Video Coding (AVC) [29] and High Efficiency Video Coding (HEVC) [30], it is necessary to develop RDH technique in videos.

Recently, the technology of RDH has been used in video privacy region protection [31] and video error concealment [32–34]. Video privacy region protection is that the privacy region is protected while keeping the non-privacy region visually intact to facilitate processing. For example, Ma et al. made use of RDH to realize video privacy region protection for cloud video surveillance in [31]. Video error concealment plays a significant role in keeping good visual quality in video transmission. In [32] Chung et al. proposed a circular embedding scheme based on RDH for improving the visual quality of marked videos. For this scheme, a constructed minimal set of Quantized Discrete Cosine Transform (QDCT) coefficients is combined with Flexible Macroblock Ordering (FMO) for reducing computational load and realizing reversibility [32]. The embedded data improves the quality of videos according to their experimental verification. Compared with the scheme in [32], the schemes of [33-34] obtain much better visual quality in terms of PSNR for videos. In [33], Xu et al. do not only fully exploit the number of coefficients, which need to be modified for reversibly hiding data, but also consider many extra nonzero residual blocks produced by data hiding. Furthermore, Xu et al. take the distribution of the motion vector data and the characteristic of histogram shifting into account for the specific two dimensional RDH scheme in [34]. Consequently, the two schemes can obtain much better visual quality of videos when compared with the method of [32]. However, the embedding capacities in [32–34] are fixed at 1188 bits per intra-frame (QCIF format: 176×144 pixels). Thus, proposing a novel RDH method to embed the same information bits but using less carriers is important. What's more, video RDH can be used in other application occasions. This has motivated us to research RDH technique in this paper.

According to the work of [35], different positions in a 4×4 block (shown in Fig. 1) have various impacts caused by DH. From low frequency area to high frequency area, changing a high frequency coefficient will result in less distortion than changing a low/middle frequency coefficient. Here, the mentioned modifications on high frequency, middle frequency and low frequency coefficients are identical. Based on this analysis, $AC_{15}$ coefficient in 4×4 blocks are modified for RDH. Thus, the proposed scheme can obtain higher embedding capacity. In addition, the proposed scheme can also keep better visual quality while keeping the same embedding payloads with some related methods.

The remainder of this paper is organized as follows. In Section 2, related works are reviewed. In Section 3, the reversible video data hiding scheme is presented. Experimental results and comparisons are given in Section 4. Finally we draw the conclusions in Section 5.

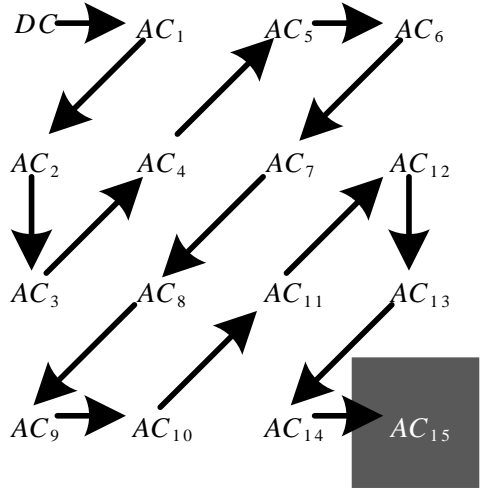

Fig. 1. QDCT coefficients in zig-zag scanning order (from low frequency to high frequency) for 4 ×4 luminance block.

## 2  Related works

Up to now, there are many RDH schemes making use of modification direction for images that have been reported. For example, Jung proposed a high-capacity RDH scheme which is based on sorting and prediction in digital images in [36]. In his scheme, each image is divided into non-overlapping 1×3 sub-blocks and each one contains three pixels. Then, the three pixels in each sub-block are sorted in ascending order. In the following, the first two pixels and the last two pixels in each sorted sub-block are used and calculated to obtain two predictor errors. Finally, two information bits can be embedded into the two predictor errors according to the histogram modification. In other words, each three pixel can embed two information bits in his RDH scheme. In our proposed scheme, however, although sometimes there may be one zero QDCT coefficient not to be paired, each zero QDCT coefficient-pair can always carry 3 bits. Additionally, Hu et al. exploited two embedding directions combined with difference expansion to propose a RDH scheme in [37]. In Hu et al.'s scheme, the inner region is used for hiding data and the outer region is shifted for correctly extracting the embedded data and completely recover the original images. For the inner region during [$-T - 1$, $T$], it will be changed to [$-2T - 2$, $2T + 1$] after the to-embedded-data is embedded into its differences. For instance, the change range of the inner region is during [-2, 1] when $T = 0$. For a zero QDCT coefficient in our proposed scheme, its change range is during [-1, 1]. Thus, the change range of Hu et al.'s scheme is greater than that of our proposed scheme that may lead to more distortion. In [38], although Li et al. make use of the modification direction for RDH, they cannot make full use of modification statuses because their scheme is based on prediction error and it need to perform the procedure of prediction two times to obtain a difference-pair ($d_1$, $d_2$). For instance, a two-dimensional difference-pair ($d_1$, $d_2$) can have four modification direction which leads to four modification statuses, i.e., ($d_1$ - 1, $d_2$), ($d_1$ + 1, $d_2$), ($d_1$ + 1, $d_2$ - 1), ($d_1$-1, $d_2$ + 1), in Li et al.'s scheme. For our scheme, a zero QDCT coefficient-pair (0, 0) can obtain eight modification statuses, i.e., (-1, 0), (0, -1), (1, 0), (0, 1), (-1, 1), (1, -1), (1, 1) and (-1, -1). Namely, a zero QDCT coefficient-pair can carry 3 bits in our proposed scheme, but a two-dimensional difference-pair can only carry 2 bits for Li et al.' s scheme. Totally, compared with those previous methods in [36–38] the proposed work make full use of not only the modification direction but also the modification statuses. And thus our proposed scheme provides more embedding payload. More details of our proposed scheme will be addressed in the following section.

## 3 Proposed scheme

This Section contains four Subsections, including zero coefficient-pair, data embedding, Data extraction and video recovery and an example of the proposed scheme. More details are addressed in the following.

3.1 Zero coefficient-pair

This subsection defines zero coefficient-pair. Before defining zero coefficient-pair, H.264/AVC video compression standard [29] is briefly reviewed. H.264/AVC uses a macroblock as the operation unit to compress videos, which are composed of many frames. In addition, these frames consist of many macroblocks. For instance, each frame of videos with the resolution of 176×144 contains 99 macroblocks, whose luminance component are sized 16×16. Namely, there are 16 blocks sized 4×4 in the luminance component of a macroblock. The operations of DCT transform and quantization in H.264/AVC [29] regard each 4×4 block as an operation unit.

After the operations of DCT transform and quantization, the coefficients in each 4×4 block are zig-zag scanned and rearranged as a vector. The scanning order is shown as Fig. 1. Without loss of generality, we assume one macroblock has four 4×4 blocks. In this case, four vectors formed from the four 4×4 blocks are taken as an example for better understanding the definition of zero coefficient-pair and they are given as Fig. 2. Since the last QDCT coefficient of each 4×4 block, i.e., $AC_{15}$ (shown as Fig. 1), has least impact on visual quality of marked videos when compared with other coefficients. In our scheme, therefore, the last QDCT coefficient of each 4×4 block may be modified for data embedding.

As shown in Fig. 2, each vector corresponds to one 4×4 block. For the last QDCT coefficient of the first vector, it is changed to vacate room for reversibility because its value is nonzero. At the data-extracting side, it is not used for extracting data but it is changed for recovering the original videos. Thus, it is referred as to QDCT recovery coefficients in this paper. For the last QDCT coefficients of the second vector and the third vector, they are paired for data embedding at the data-embedding side. Namely, the two coefficients are called as a coefficient-pair. Similarly, they are exploited to extract the embedded data and thus they are named QDCT extracting coefficients in this paper. Noted that, although the last coefficient of the fourth vector is zero, but it is not exploited for data embedding since there is not additional zero QDCT coefficient to pair with it.

| first  | 19 | 0 | 0  | 8 | 8 | 4  | 0 | 0 | 0  | 0  | 0 | 0 | 0  | −1 | 0 | 1 |
| second | −7 | 0 | 0  | 0 | 0 | 3  | 0 | 0 | 0  | −1 | 1 | 0 | 1  | −1 | 0 | 0 |
| third  | 1  | 0 | −1 | 0 | 1 | 0  | 0 | −1 | 0 | −1 | 1 | 0 | 0  | 0  | 0 | 0 |
| fourth | −7 | 0 | 1  | 1 | 1 | −2 | 0 | 0 | −1 | 0  | −1 | 0 | −1 | 0  | 0 | 0 |

Fig. 2. An example of coefficient-pair definition.

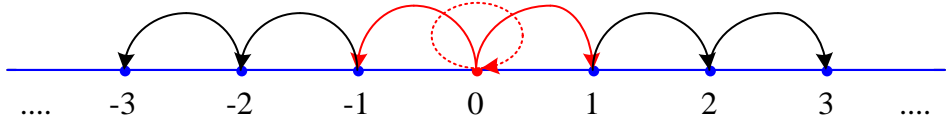

Fig. 3. Histogram modification mechanism for a zero QDCT coefficient in the proposed RDH method.

### 3.2 Data Embedding

In general, DE [19-21] and histogram shifting (HS) [19, 39] are two commonly used approaches to realize reversibility of data hiding. In the proposed RDH method, we exploit improved HS to realize the reversibility.

At the data-hider side, also called encoder side, the macroblocks with intra-frame 4×4 prediction modes in intra frames are first selected as embeddable blocks. Furthermore, since the last coefficient has a least impact on visual quality of marked videos when compared with coefficients located in other areas of the 4×4 block (shown as Fig. 1). Therefore, the last QDCT coefficient, i.e., $AC_{15}$, in each 4×4 block of the embeddable macroblocks may be selected for embedding data only if $AC_{15} = 0$ and there exists at least one nonzero QDCT coefficient in the corresponding 4×4 block. Each embeddable 4×4 block is zig-zag scanned (shown as Fig. 1) and rearranged to a vector. As shown in Fig. 2, the last coefficients of the second vector, the third vector and the fourth vector can be used to embed information bits. For one zero QDCT coefficient, it has three statuses (shown as Fig. 3) by keeping unchanged, increasing or decreasing by 1. In order to make full use of all statuses of one zero coefficient, we use zero coefficient-pair to improve the embedding capacity and the zero coefficient-pair has been described in Subsection 3.1. One zero coefficient-pair can be embedded 3 information bits in the proposed scheme.

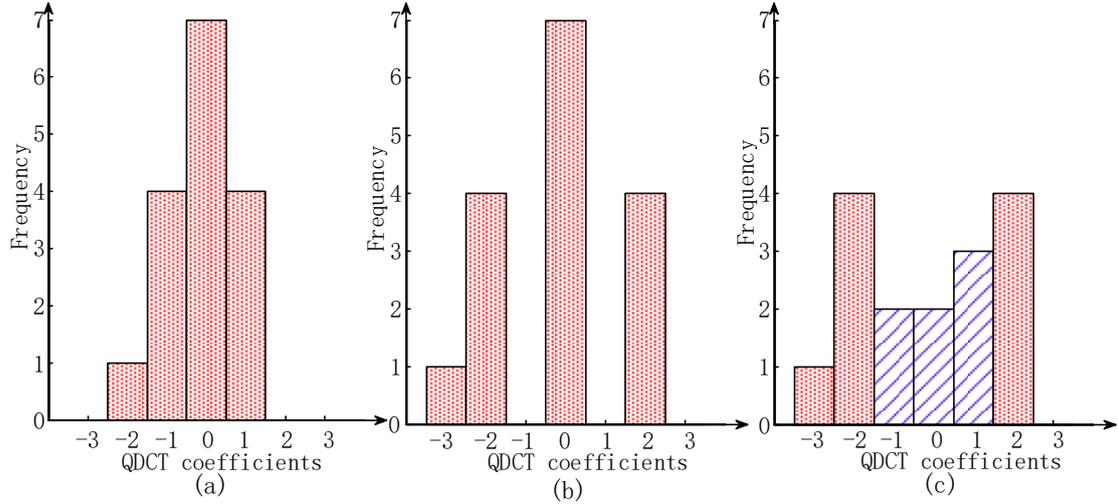

Fig. 4. A distribution of all coefficients $AC_{15}$ in a macroblock. (a) A set of all coefficients $AC_{15}$. (b) Shift positive and negative QDCT coefficients to the right position and the left position, respectively. (c) After embedding secrete data into zero QDCT coefficient pairs.

The embedding procedure of RDH is addressed here in detail. In one embeddable macroblock, $AC_{15}$, i.e., the last QDCT coefficient of one 4×4 block (like the first vector in Fig. 2), will be changed by

$$AC' = \begin{cases} AC_{15}+1, & AC_{15} \geq 0 \\ AC_{15}-1, & AC_{15} \leq 0 \end{cases} \quad (1)$$

when $AC_{15} \neq 0$. In the following, to embed data into the video the zero coefficient-pairs will be changed by

$$C = \begin{cases} (1,1), & if\ B = '000' \\ (-1,0), & if\ B = '001' \\ (-1,1), & if\ B = '010' \\ (0,-1), & if\ B = '011' \\ (0,0), & if\ B = '100' \\ (0,1), & if\ B = '101' \\ (1,-1), & if\ B = '110' \\ (1,0), & if\ B = '111' \\ (-1,-1), & not\ used \end{cases} \quad (2)$$

where $C = (c_1, c_2)$ represents a modified QDCT coefficient-pair and $B = "b_1b_2b_3"$ denotes the to-be-embedded three information bits. Noted that, the modified QDCT coefficient-pair is obtained by changing the original zero QDCT coefficient-pair to embed data. By doing this in an embeddable macroblock, the distribution of all $AC_{15}$ coefficients will change. To better understand this, we exploit Fig. 4 to explain it.

Fig. 4(a) denotes the original distribution of all $AC_{15}$ coefficients in an embeddable macroblock and Fig. 4(b) is its correspondingly modified distribution by Eq. (1). Obviously, there are 7 zero QDCT coefficients that can be used to embed data. Since zero coefficient-pair is exploited in our proposed scheme for embedding data. Therefore, in this macroblock three zero coefficient-pairs are made use of embedding data and the rest one zero coefficient is kept unchanged. Assume we embed information bits "111 011 111" into the three zero coefficient-pairs according to Eq. (2). Finally, the distribution is modified from Fig. 4(b) to Fig. 4(c).

3.3 Data extraction and video recovery

Similarly, at the data extractor side, also called the decoder side, the macroblocks with intra-frame 4×4 prediction modes in intra frames are first selected as embeddable blocks. Moreover, in each embeddable macroblock, a 4×4 block is selected as an embeddable block when it has at least one zero QDCT coefficient. Likewise, each embeddable 4×4 block is zig-zag scanned (shown as Fig. 1) and rearranged to a vector. In each vector, if the absolute value of $AC_{15}$ is greater than 1, it is referred as to QDCT recovery coefficient. In order to restore the original compression videos, this type coefficient will be changed by

$$AC' = \begin{cases} AC_{15} - 1, AC_{15} > 1 \\ AC_{15} + 1, AC_{15} < -1 \end{cases} \quad (3)$$

In addition, if the absolute value of $AC_{15}$ in each vector is less than 1, it is called QDCT extracting coefficient. In each macroblock, all QDCT extracting coefficients are paired two by two, named coefficient-pair. If there is still one QDCT extracting coefficient is not paired, it means this coefficient is not changed at the data-hider side. Thus, at the data-extractor side, we do not need to change this coefficient for restoring the original compression videos. For each coefficient-pair, the embedded information is extracted by

$$B = \begin{cases} '000', if\ C = (1,1) \\ '001', if\ C = (-1,0) \\ '010', if\ C = (-1,1) \\ '011', if\ C = (0,-1) \\ '100', if\ C = (0,0) \\ '101', if\ C = (0,1) \\ '110', if\ C = (1,-1) \\ '111', if\ C = (1,0) \end{cases} \quad (4)$$

where $B$ = "$b_1b_2b_3$" denotes the extracted information bits and $C = (c_1, c_2)$ represents the coefficient-pair. After the embedded information is extracted out in the current macroblock, all coefficient-pairs are set to 0. For an embeddable macroblock, the changed QDCT coefficients are recovered to their original values corresponding to the coder side. Meanwhile, the distribution of all $AC_{15}$ in this embeddable macroblock is recovered. Like Fig. 4, the distribution is restored from Fig. 4(c) to Fig. 4(a). By doing this in each embeddable macroblock, all the embedded information is extracted out and the original compression video is obtained.

| first  | 19 | 0 | 0  | 8 | 8 | 4  | 0 | 0 | 0  | 0  | 0  | 0 | 0  | -1 | 0 | 2  |
|--------|----|---|----|---|---|----|---|---|----|----|----|---|----|----|---|----|
| second | -7 | 0 | 0  | 0 | 0 | 3  | 0 | 0 | 0  | -1 | 1  | 0 | 1  | -1 | 0 | 1  |
| third  | 1  | 0 | -1 | 0 | 1 | 0  | 0 | -1| 0  | -1 | 1  | 0 | 0  | 0  | 0 | -1 |
| fourth | -7 | 0 | 1  | 1 | 1 | -2 | 0 | 0 | -1 | 0  | -1 | 0 | -1 | 0  | 0 | 0  |

Fig. 5. An example to data embedding corresponding to Fig. 2.

3.4 An example of the proposed scheme

In this Subsection, we will take an example to address the proposed scheme for better understanding. Without loss of generality, assume Fig. 2 shows four vectors corresponding to four embeddable 4×4 blocks in an embeddable macroblock.

Here, assume the to-be-embedded information bits are "110". Using the proposed method, according to Eq. (1) the $AC_{15}$ coefficient of the first vector is changed to 2. This procedure corresponds to the procedure from Fig. 4(a) to Fig. 4(b). The zero coefficient-pair, which is made of the last zero coefficient of the second vector and the third vector, is changed to embed "110" by Eq. (2). Therefore, Fig. 5 can be obtained. Here, the procedure corresponds to the procedure from Fig. 4(b) to Fig. 4(c). During data extraction, the last coefficients of the second vector and the third vector are used to extract data by Eq. (4) and then they are set to 0. This procedure corresponds to the procedure from Fig. 4(c) to Fig. 4(b). In the following, the coefficients, whose absolute value are greater than 1, are changed by Eq. (3). Like the first vector in Fig. 5, its last coefficient is restored to 1. Finally, Fig. 5 becomes Fig. 2 after data extraction and coefficient recovery.

## 4 Experimental results and analysis

Table 1 Configuration parameters of the JM12.0 software

| Parameter | Configuration |
|---|---|
| Profile | Main |
| Frame Rate | 15 |
| Rate Distortion Optimization | On |
| IntraPeriod | 15 |
| Symbol mode | 0: CAVLC |
| FrameSkip | 1 |
| NumberBFrame | 1 |
| QP | 28 |

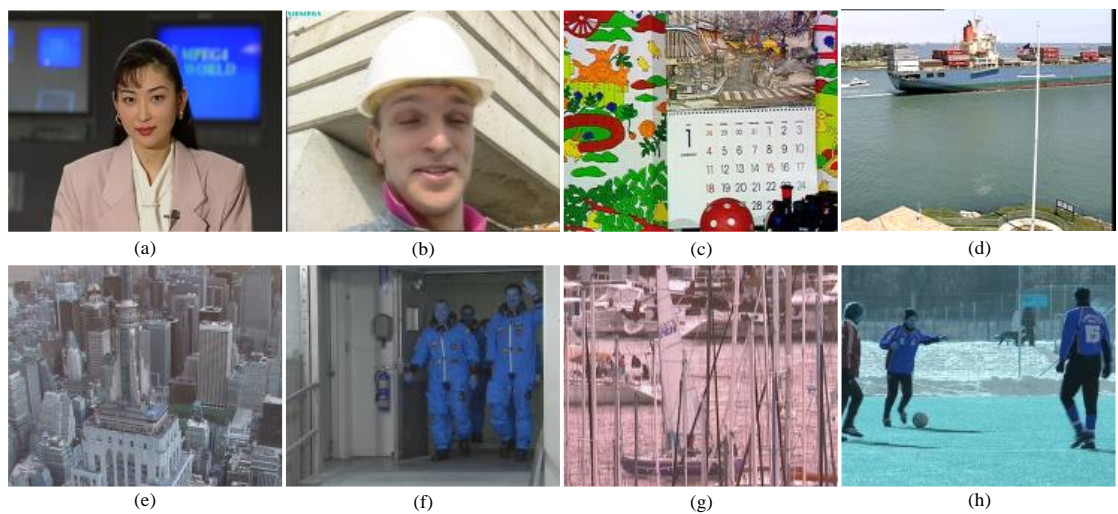

Fig. 6. Test video with resolution of 352×288. (a) Akiyo. (b) Foreman. (c) Mobile. (d) Container. (e) City. (f) Crew. (h) Soccer.

The proposed method has been implemented in the H.264/AVC reference software JM12.0 [40]. To evaluate the performance of the proposed method objectively, the 8 video sequences, i.e., Akiyo, Foreman, Mobile, Container, City, Crew, Harbour and Soccer, are used to be test samples (shown as Fig. 6). Each video contains two kinds of resolution (also called as Frame Size in this paper) which are shown in Tables 2-4 and the parameter RateControlEnable is set to "0". In addition, some configuration parameters are given in Table 1 and other configuration parameters, which are not referred to in this paper, retain their default values. The standard video sequences are encoded into 300 frames at 15 frames per second. For the encoded video sequences with 300 frames, there are only 10 intra frames between them. The group of picture (GOP) is IBPBPBPBPBPBPBPBPBPBPBP. The performance comparison between the schemes [15, 18, 27, 28] and the proposed scheme is illustrated in subsections 4.2-4.4. It is noted that the value of Threshold in [18] is set to 10 and the two coefficient-pairs {$\tilde{Y}_{30}$, $\tilde{Y}_{32}$} and {$\tilde{Y}_{23}$, $\tilde{Y}_{03}$} are exploited to hide data when $|\tilde{Y}_{00}|>0$ in [15]. For the method in [28], macroblocks with intra frame 4×4 prediction modes in intra frames are selected as embeddable blocks and furthermore 4×4 blocks of embeddable macroblocks are selected as embeddable if the absolute value of the first nonzero coefficients in 4×4 blocks is less than 2.

4.1 Relationship between coefficient-pairs and embedding capacity

In the proposed method, all $AC_{15}$ coefficients are classified into three types, including zero coefficient-pairs, unpaired zero coefficients and nonzero coefficients. The zero coefficient-pairs are used for data embedding. However, the unpaired zero coefficients and nonzero coefficients are unexploited for data hiding. In order to address the relationship between the zero coefficients constituting the zero coefficient-pairs and embedding capacity, we draw Fig. 7 and analyze it as follows.

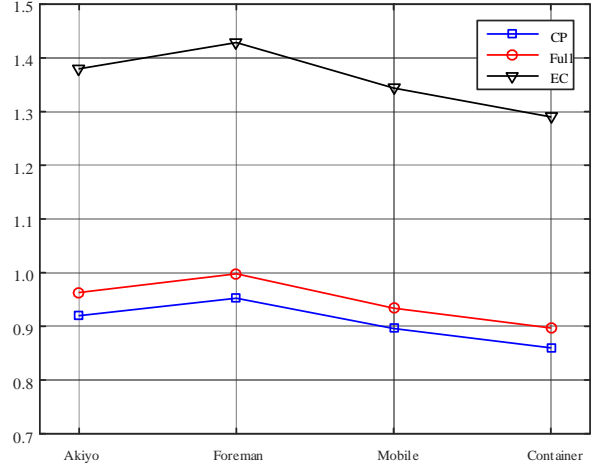

Fig. 7. Relationship between the number of zero $AC_{15}$ coefficients and the embedding capacity (resolution: 176×144, QP:28).

As shown in Fig. 7, we draw three curves, i.e., *CP*, *Full* and *EC*. *CP* is defined by

$$CP = \frac{the\ number\ of\ zero\ AC_{15}\ coefficients\ used\ for\ embedding\ data}{the\ number\ of\ AC_{15}\ coefficients\ in\ embeddable\ 4\times 4\ blocks} \times 100\% \quad (5)$$

Likewise, *Full* and *EC* can be defined. *Full* denotes the rate of the number of all zero $AC_{15}$ coefficients and the number of $AC_{15}$ coefficients in embeddable 4×4 blocks. *EC* represents the embedding rate, i.e., the rate of the number of the embedded information bits and the number of $AC_{15}$ coefficients in embeddable 4×4 blocks. For different videos, *EC* mainly depends on *CP* or *Full* (shown as Fig. 7). The larger *CP* or *Full* is, the larger *EC* is. In the most ideal case, all $AC_{15}$ coefficients are zero and they are paired for data embedding. In other words, *EC* is 1.5. In practice, it is too difficult to meet this. Noted that, *EC* is during [1.25, 1.45] and *CP* and *Full* are during [0.85, 1]. That is to say most $AC_{15}$ coefficients are 0 in practice. Furthermore, most zero $AC_{15}$ coefficients can be paired for data hiding. Therefore, the more the zero coefficient-pairs are, the higher the embedding capacity is.

### 4.2 Embedding capacity

Table 2 compares the maximum embedding capacity (also referred to as embedding payloads) of the proposed scheme and four previous schemes [15, 18, 27, 28]. The schemes in [15, 18] are irreversible but the rest two scheme in [27, 28] are reversible. As shown in Table 2, for each test video, the proposed method can obtain the highest embedding capacity when compared with the four methods. For instance, the test video City with the resolution of 352×288 provides embedding capacity of 71823 bits when using the proposed scheme. At the same time, the four schemes can respectively obtain 7731, 14867, 11467 and 34161 bits on City. Obviously, according to Table 2 the average maximum embedding capacity of the proposed method is considerably higher than these of the four previous methods. In summary, the proposed method has indeed advantages in embedding capacity.

Table 2 Comparison of maximum embedding capacity (bits) of the proposed scheme and four previous schemes [15, 18, 27, 28] (QP: 28).

| Sequences | Frame Size | No. of Frame | Embedding capacity | | | | |
|---|---|---|---|---|---|---|---|
| | | | [18] | [15] | [27] | [28] | Proposed |
| Akyio | 176×144 | 300 | 1235 | 2213 | 1459 | 4499 | 9579 |
| | 352×288 | 300 | 2395 | 5820 | 2942 | 11045 | 24195 |
| Foreman | 176×144 | 300 | 2108 | 2705 | 1618 | 7604 | 15057 |
| | 352×288 | 300 | 5982 | 8178 | 5008 | 17611 | 47271 |
| Mobile | 176×144 | 300 | 9904 | 4263 | 1722 | 3462 | 18357 |
| | 352×288 | 300 | 32338 | 15489 | 7179 | 14166 | 72333 |
| Container | 176×144 | 300 | 2742 | 2362 | 1367 | 2453 | 9792 |
| | 352×288 | 300 | 8178 | 9127 | 5191 | 13430 | 43185 |
| City | 352×288 | 300 | 7731 | 14867 | 11467 | 34161 | 71823 |
| | 704×576 | 300 | 9700 | 56058 | 44822 | 153601 | 268119 |
| Crew | 352×288 | 300 | 937 | 4454 | 2806 | 20091 | 21822 |
| | 704×576 | 300 | 1515 | 26799 | 15206 | 54115 | 129504 |
| Harbour | 352×288 | 300 | 11247 | 22067 | 14672 | 55311 | 86973 |
| | 704×576 | 300 | 2287 | 78385 | 45569 | 283772 | 310032 |
| Soccer | 352×288 | 300 | 3806 | 10165 | 6435 | 22517 | 52956 |
| | 704×576 | 300 | 4076 | 40409 | 28112 | 102586 | 205332 |
| Average | - | - | 6636 | 18960 | 12223 | 50026 | 86645 |

4.3 Visual quality

In this Subsection, we first use Fig. 8 to evaluate the visual quality of marked videos subjectively. As shown in Fig. 8, although the methods of DH and RDH certainly result in video distortion, there are not significant distortion are observed. In addition, to measure the visual quality of marked videos objectively, Peak-Signal-Noise-Ratio (PSRN) is used to compare the transparency [41] between the proposed method and the methods of [15, 18, 27, 28]. PSNR values of these methods are given Table 3. Noted that, Table 3 corresponds to Table 2. In Table 2, "PSNR1" is calculated between the original frames and the compressed/decompressed frames. Similarly, "PSNR2" is calculated between the original frames and the compressed/decompressed frames with the embedded data. Usually, the methods of DH have better visual quality when compared with the methods of RDH. Meanwhile, DH methods or RDH methods provide the embedding capacity that also has an important impact on visual quality. As seen in Table 3, using the methods of [15, 18] leads to better visual quality in terms of PSNR when compared with using the proposed scheme and the method in [28]. Although the method of [27] provides better visual quality than the methods of [15, 18], its embedding capacity is not higher than that of [28] and the proposed method. When comparing the RDH methods, our proposed method has higher PSNR than the method of [28]. Although our proposed scheme do not have better visual quality than the method of [27], our proposed scheme provides far higher embedding capacity than the method of [27]. In addition, the method in [27]

Table 3 Comparison of visual quality in terms of PSNR (dB) between the schemes (QP: 28)

| Sequences | Frame Size | No. of Frame | PSNR1 Original | PSNR2 [18] | PSNR2 [15] | PSNR2 [27] | PSNR2 [28] | PSNR2 Proposed |
|---|---|---|---|---|---|---|---|---|
| Akyio | 176×144 | 300 | 39.32 | 37.71 | 38.64 | 39.01 | 33.58 | 35.81 |
|  | 352×288 | 300 | 40.71 | 39.43 | 40.12 | 40.54 | 33.85 | 36.73 |
| Foreman | 176×144 | 300 | 36.81 | 34.21 | 36.61 | 36.79 | 31.75 | 32.80 |
|  | 352×288 | 300 | 36.90 | 34.50 | 36.89 | 37.03 | 31.65 | 32.67 |
| Mobile | 176×144 | 300 | 34.22 | 30.50 | 34.00 | 34.12 | 31.99 | 32.46 |
|  | 352×288 | 300 | 35.01 | 31.77 | 34.79 | 34.90 | 31.88 | 32.55 |
| Container | 176×144 | 300 | 36.90 | 34.72 | 36.36 | 36.63 | 31.44 | 33.66 |
|  | 352×288 | 300 | 36.75 | 34.24 | 36.26 | 36.53 | 29.13 | 32.78 |
| City | 352×288 | 300 | 35.77 | 33.80 | 35.57 | 35.64 | 27.43 | 30.87 |
|  | 704×576 | 300 | 35.89 | 34.87 | 35.72 | 35.76 | 27.02 | 28.22 |
| Crew | 352×288 | 300 | 37.90 | 37.74 | 37.75 | 37.87 | 30.27 | 35.54 |
|  | 704×576 | 300 | 38.01 | 37.93 | 37.94 | 38.06 | 30.65 | 34.16 |
| Harbour | 352×288 | 300 | 35.14 | 32.61 | 34.83 | 35.03 | 25.06 | 29.39 |
|  | 704×576 | 300 | 36.12 | 35.21 | 35.84 | 36.06 | 23.72 | 27.72 |
| Soccer | 352×288 | 300 | 36.63 | 35.53 | 36.61 | 36.69 | 29.65 | 32.14 |
|  | 704×576 | 300 | 37.11 | 36.92 | 37.14 | 37.19 | 27.31 | 32.18 |
| Average | - | - | 37.02 | 35.11 | 36.57 | 36.74 | 29.78 | 32.48 |

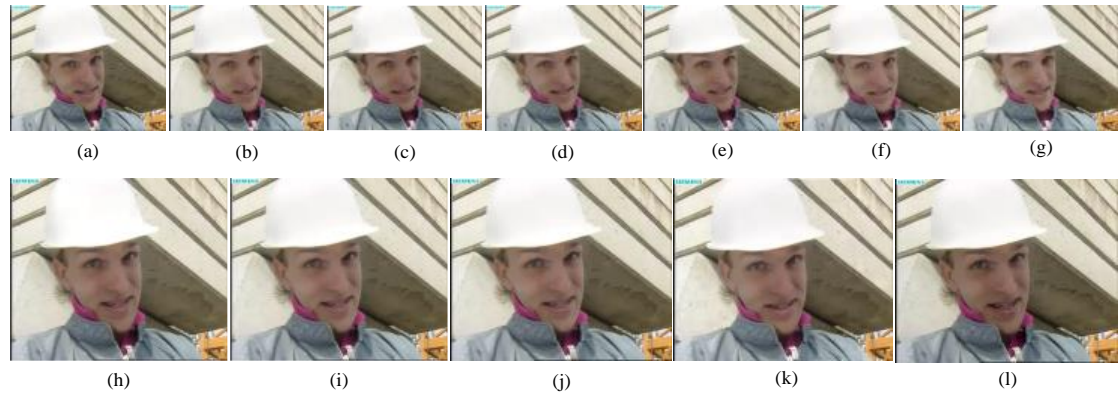

(a) (b) (c) (d) (e) (f) (g)

(h) (i) (j) (k) (l)

Fig. 8. Original, compressed/decompressed and marked frames (150th frame of Foreman). (a-g) are with the size of 176×144 and (h-l) are with the size of 352 ×288. (a) is the original frame. (b) is the compressed/decompressed frame. (c) and (h) are embedded additional data into by the scheme in [18]. (d) and (i) are embedded additional data into by the scheme in [15]. (e) and (j) are embedded additional data into by the scheme in [27]. (f) and (k) are embedded additional data into by the scheme in [28]. (g) and (l) are embedded additional data into by the proposed scheme.

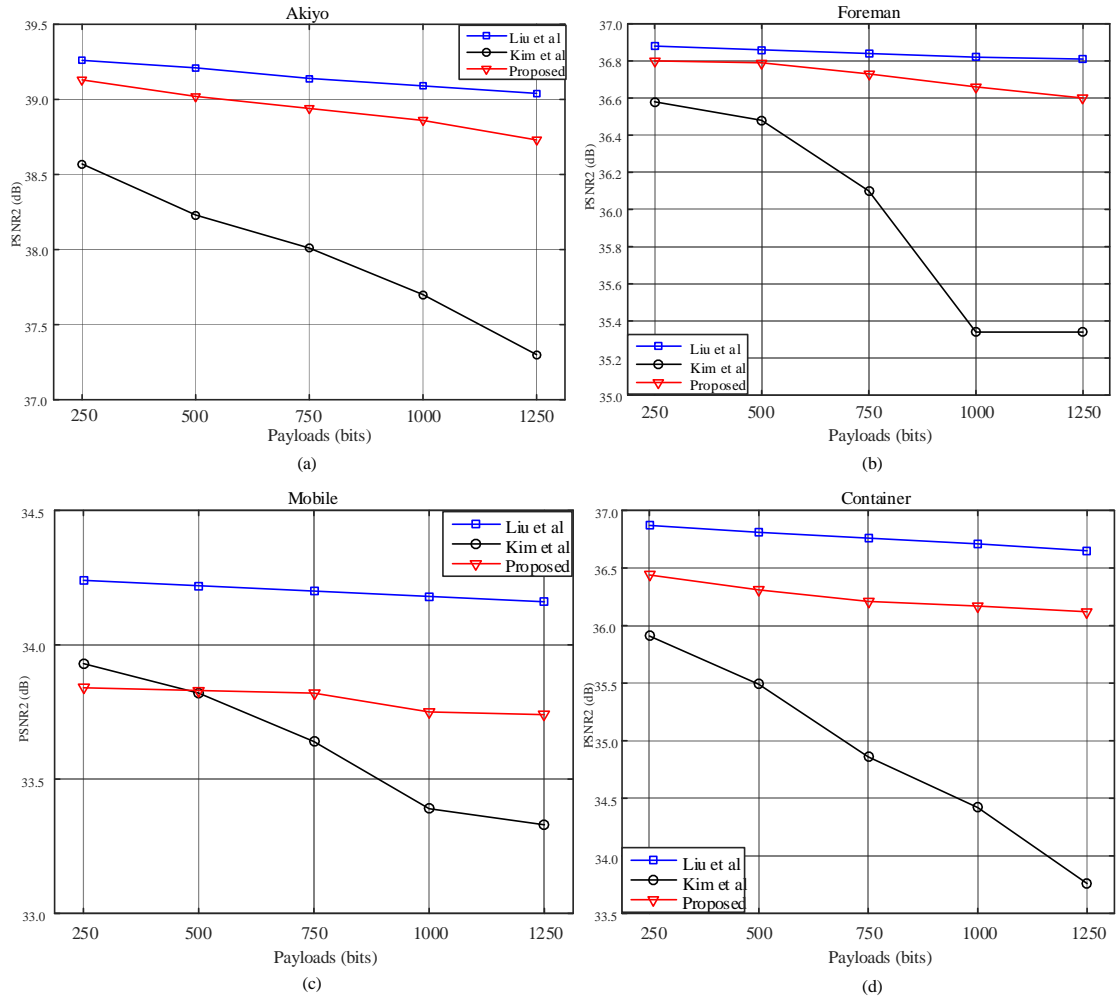

Fig. 9 Relationship between visual quality (PSNR: dB) and embedding payloads (bits) of the proposed scheme and two schemes [27, 28] on four videos with the resolution of 176×144. (a) Akiyo. (b) Foreman. (c) Mobile. (d) Container.

prevents intra frame distortion from spreading, thus it can obtain better visual quality.

    For a better comparison, we fix the embedding payloads at 250, 500, 750, 1000 and 1250 bits on Akiyo, Foreman, Mobile and Container, which are with the resolution of 176×144. As seen in Fig. 9, on each video Liu et al.'s scheme [27] has a less impact on visual quality of PSNR. However, Kim et al.'s scheme [28] provides the largest impact on visual quality when compared with the proposed scheme and the scheme of [27]. In addition, using the proposed scheme and Liu et al.'s scheme has a close degradation of PSNR. In other words, using the two schemes leads to a very close difference between the maximum PSNR value and the minimum PSNR value (shown in Fig. 9). However, exploiting Kim et al.'s scheme results in far larger difference of that than exploiting the proposed scheme and the scheme of [27]. The scheme in [27] can avoid intra frame distortion drift, thus resulting in best visual quality compared with the rest two schemes. For our proposed method, the last nonzero coefficients of each embeddable 4×4 blocks, i.e., $AC_{15}$, may be changed for DH and they are usually located in high

Table 4 Comparison of bit rate variation (kbps) between the methods (QP: 28)

| Sequences | Frame Size | No. of Frame | Bit rate Original | Bit rate with embedding data | | | | |
|---|---|---|---|---|---|---|---|---|
| | | | | [18] | [15] | [27] | [28] | Proposed |
| Akyio | 176×144 | 300 | 19.66 | 19.78 | 20.30 | 19.86 | 20.30 | 22.06 |
| | 352×288 | 300 | 56.84 | 57.10 | 58.46 | 57.26 | 58.32 | 62.94 |
| Foreman | 176×144 | 300 | 70.72 | 71.09 | 71.67 | 70.93 | 72.03 | 74.34 |
| | 352×288 | 300 | 236.98 | 237.65 | 238.76 | 237.67 | 239.41 | 248.17 |
| Mobile | 176×144 | 300 | 182.67 | 183.88 | 183.19 | 183.88 | 182.79 | 185.62 |
| | 352×288 | 300 | 771.38 | 774.00 | 774.01 | 772.28 | 775.10 | 783.76 |
| Container | 176×144 | 300 | 26.80 | 27.11 | 27.26 | 26.99 | 27.17 | 28.62 |
| | 352×288 | 300 | 122.11 | 122.50 | 123.99 | 122.78 | 123.76 | 130.53 |
| City | 352×288 | 300 | 213.47 | 214.82 | 217.14 | 214.95 | 218.46 | 228.64 |
| | 704×576 | 300 | 1009.29 | 1010.93 | 1024.25 | 1015.06 | 1029.67 | 1069.27 |
| Crew | 352×288 | 300 | 462.49 | 462.63 | 465.46 | 463.31 | 465.30 | 474.38 |
| | 704×576 | 300 | 1418.39 | 1417.76 | 1427.82 | 1420.09 | 1425.05 | 1456.40 |
| Harbour | 352×288 | 300 | 716.87 | 717.88 | 721.46 | 718.65 | 724.73 | 732.14 |
| | 704×576 | 300 | 2559.47 | 2558.57 | 2580.65 | 2565.12 | 2593.15 | 2630.54 |
| Soccer | 352×288 | 300 | 380.94 | 381.38 | 383.62 | 381.12 | 383.82 | 393.32 |
| | 704×576 | 300 | 1306.76 | 1309.91 | 1321.42 | 1313.85 | 1319.77 | 1354.66 |
| Average | - | - | 597.18 | 597.94 | 602.47 | 598.99 | 603.68 | 617.21 |

frequency area of 4×4 blocks. For the method of [28], the modified coefficients mainly located in middle frequency area of 4×4 blocks. Therefore, our proposed method outperforms the method of [28] in terms of PSNR when embedding payloads are the same.

4.4 Bit-rate variation

Table 4 give original bit-rates caused by H.264/AVC encoder and marked bit-rates caused by data hiding H.264/AVC encoder. Noted that, Table 4 corresponds to Tables 2 and 3. According to Table 4, the methods of DH and RDH result in bit-rate increase. For instance, the average original bit-rate is about 597.18 kbps and the average marked bit-rates are 597.94, 602.47, 598.99, 603.68 and 617.21 kbps, respectively. To further evaluate bit-rate variation, we first define Bit-rate Increment Ratio (BIR) by

$$BIR = \frac{BR(Marked) - BR(Original)}{BR(Original)} \times 100\% \quad (6)$$

where BR(Marked) is the bit-rate generated by data hiding encoder, and BR(Original) is the bit-rate generated by the original encoder. Based on this, Fig. 10 is given to measure the relationship between embedding payloads and BIRs. Likewise, embedding payloads are fixed at 250, 500, 750, 1000 and 1250 bits.

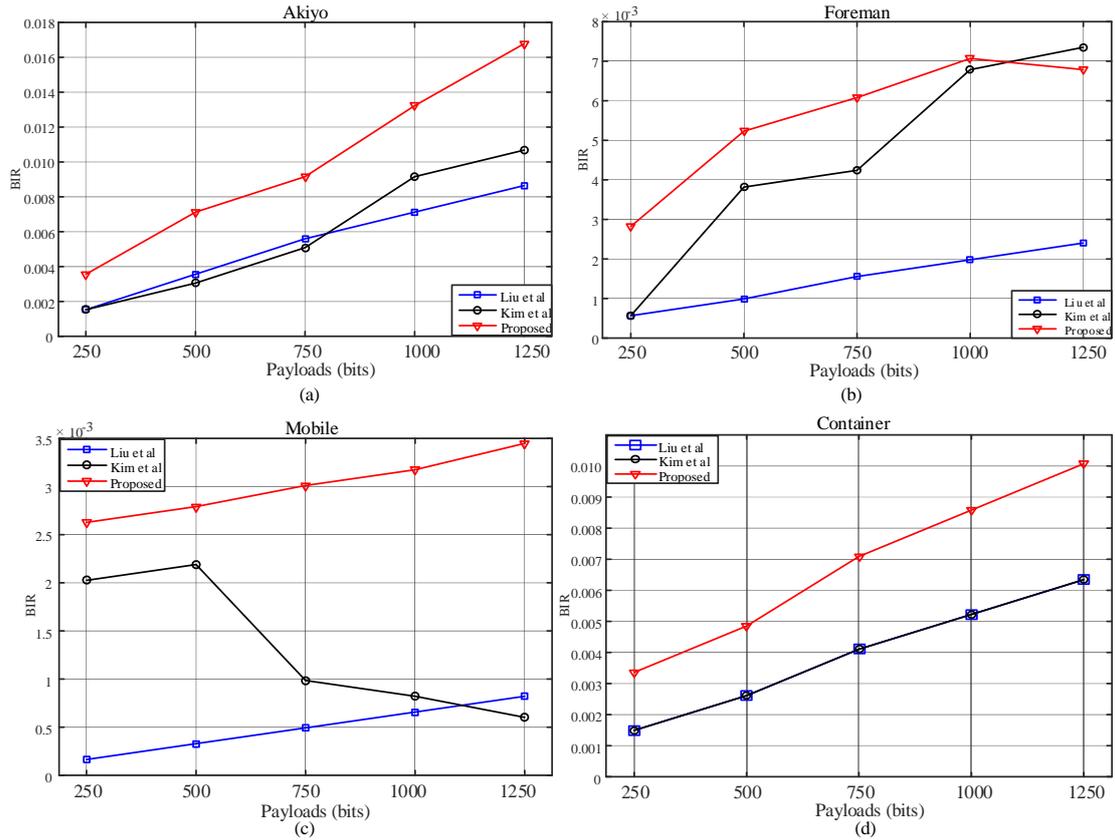

Fig. 10 Relationship between embedding payloads and BIRs of the proposed method and the methods of [27, 28] on four videos with the resolution of 176×144. (a) Akiyo. (b) Foreman. (c) Mobile. (d) Container.

As seen in Fig. 10, on the four videos RDH methods have led to bit-rate increase. According to subFigs. 10(a)-(d), our proposed method obtain the highest BIR values at corresponding embedding payloads. In most cases, zero coefficients, i.e., zero coefficient-pairs, may be changed to nonzero coefficients in our proposed scheme. Furthermore, according to H.264/AVC [29] it needs more coding and thus significantly making bit-rate increase. For Liu et al.'s scheme, only a part of nonzero coefficients are changed that leads to less increase in bit-rate. As for Kim et al.'s method, the change of coefficients is complex. It is possible that nonzero coefficient is modified to zero coefficient and vice versa. For example, most nonzero coefficients are changed to zero coefficients that may result in Fig. 10(c). On the contrary, Fig. 10(b) will happen. Otherwise, it will lead to Figs. 10(a) and 10(d). Actually, all the three RDH methods provide less BIR increase. The most significant BIR increase is about 0.017 in Fig. 10(a).

Totally, our proposed method has indeed advantages in embedding capacity when compared with the methods of [27, 28]. For PSNR values at the same embedding payloads, although our proposed method is not as good as Liu et al.'s method, our proposed method outperforms Kim et al.'s scheme. In the future, we think avoiding intra frame distortion drift and defining distortion cost of elements can be exploited for RDH. That may make RDH methods produce better performance.

## 5 Conclusions

In this paper, we take advantage of zero coefficients to propose a novel RDH method, which is based on H.264/AVC compression standard. The proposed method exploits each zero coefficient-pair to embed 3 information bits that means embedding ratio is 1.5. Therefore, the proposed RDH method can provide a higher embedding capacity. In addition, our proposed scheme outperforms some previous RDH methods with intra frame distortion drift in terms of PSNR when embedding payloads are set to the same. In the future, we think stopping intra frame distortion drift and defining distortion cost for each element may be exploited in RDH technology. In addition, more coding parameters may be considered for RDH. These will be a research direction of RDH and also result in far better performance in terms of visual quality and compression efficiency when keeping the same embedding payloads.

## Acknowledgement

The authors would like to thank the reviewers for their insightful comments and helpful suggestions.This work was supported by the National Natural Science Foundation of China ( NSFC) under the grant No. U1536110 and Doctoral Innovation Fund Program of Southwest Jiaotong University.